# Temperature Evolution of Itinerant Ferromagnetism in SrRuO$_3$ Probed by Optical Spectroscopy

D. W. Jeong,[1,2] Hong Chul Choi,[3] Choong H. Kim,[2] Seo Hyoung Chang,[1,2] C. H. Sohn,[1,2] H. J. Park,[1,2] T. D. Kang,[1,2] Deok-Yong Cho,[1,2] S.H. Baek,[4] C. B. Eom,[4] J. H. Shim,[3,5] J. Yu,[2] K. W. Kim,[6] S. J. Moon,[7†] and T. W. Noh[1,2,*]

[1] *Center for Functional Interfaces of Correlated Electron Systems, Institute for Basic Science (IBS), Seoul 151-747, Korea*

[2] *Department of Physics and Astronomy, Seoul National University, Seoul 151-747, Korea*

[3] *Department of Chemistry, Pohang University of Science and Technology, Pohang 790-784, Korea*

[4] *Department of Materials Science and Engineering, University of Wisconsin–Madison, Madison, Wisconsin 53706, USA*

[5] *Division of Advanced Nuclear Engineering, Pohang University of Science and Technology, Pohang 790-784, Korea*

[6] *Department of Physics, Chungbuk National University, Cheongju 361-763, Korea*

[7] *Department of Physics, Hanyang University, Seoul 133-791, Korea*

**Abstract**

The temperature ($T$) dependence of the optical conductivity spectra, $\sigma(\omega)$, of single crystal SrRuO$_3$ thin film is studied over a $T$ range from 5 to 450 K. We observed significant $T$ dependence of the spectral weights of the charge transfer and interband $d$-$d$ transitions across the ferromagnetic Curie temperature ($T_c \sim 150$K). Such $T$-dependence was attributed to the increase in the Ru spin moment, which is consistent with the results of density-functional-theory calculations. $T$ scans of $\sigma(\Omega, T)$ at fixed frequencies $\Omega$ reveal a clear $T^2$ dependence below $T_c$, demonstrating that the Stoner mechanism is involved in the evolution of the electronic structure. In addition, $\sigma(\Omega, T)$ continues to evolve at temperatures above $T_c$, indicating that the local spin moment persists in the paramagnetic state. This suggests that SrRuO$_3$ is an intriguing oxide system with itinerant ferromagnetism.

One of the central questions in condensed matter physics is how the electronic correlation can induce magnetic orders in metallic systems [1-4]. Several decades ago, Stoner pointed out that the repulsive Coulomb interaction ($U$) among itinerant electrons can induce a spin split electronic band structure, resulting in the preference of spin direction [5]. The Stoner model successfully describes non-integer magnetic moments in metallic ferromagnets, such as Fe, Co, and Ni, which cannot be explained by the Heisenberg-type local spin–spin interactions [1, 3]. The ground-state electronic structures have also been described successfully by Stoner exchange splitting between the majority and minority spin bands [1, 6]. Therefore, the Stoner theory is one of the widely accepted models to describe itinerant ferromagnetism (FM).

$SrRuO_3$ is a metallic oxide with a moderate ferromagnetic Curie temperature ($T_c$) of 160K [7]. Despite extensive studies [8-14], there has been long controversy on whether FM originates from local moments or itinerant electrons. The saturation magnetic moment of $SrRuO_3$ is $1.6\mu_B$ [7], which is a non-integer value. Its low-temperature electronic structure can be explained quite well, with band-structure calculations that include Stoner exchange splitting [10, 11]. Therefore, $SrRuO_3$ has been considered as the first example of an oxide showing itinerant FM. However, contrary to the predictions of the Stoner model, magneto-optic Kerr effect data did not show a significant band shift up to $\sim T_c$ [8]. Recent Angle-resolved photoemission spectroscopy (ARPES) studies did not reveal any significant energy shift in the exchange split band, up to $\sim T_c$ [9]. These studies concluded that the local moment FM was more appropriate for describing $SrRuO_3$ [8, 9]. Such a controversy has raised a fundamental question whether itinerant ferromagnetism can develop (or be realized) at all in oxide materials that have rather narrow bands and rather strong electronic correlation. Solving this long standing controversy is also important from application as well as scientific points of view. Because $SrRuO_3$ has been very widely used for oxide electronic applications, such as magnetic tunnel junctions and spintronic devices [15-19], it is quite important to elucidate the nature of FM in this intriguing material.

In this Letter, we report the electronic structure of SrRuO$_3$ thin films over a temperature ($T$) range of 5 to 450K using optical spectroscopy. Compared to ARPES, our spectroscopic technique is much more sensitive to $T$-dependent spectral weight changes. We found that our observed $T$-dependent optical conductivity spectra $\sigma(\omega)$ can be explained by the local spin density approximation (LSDA) calculations in terms of variation in the spin polarization. $T$-scan measurements of $\sigma(\Omega, T)$ at fixed frequencies ($\Omega$) further indicated $T^2$-scaling behavior below $T_c$, which is reminiscent of the $T^2$-dependence of spin polarization in the Stoner theory. However, there is an optical signature that suggests the existence of persistent local spin-split bands even in the paramagnetic state with $T > T_c$. These experimental results agree with the predictions of fluctuating local band theory of itinerant FM, in which the band structure itself is determined locally by the Stoner mechanism, but its average magnetization direction vanishes at $T_c$ [4].

An epitaxial SrRuO$_3$ film, with a thickness of 5000- Å, was deposited on a (100) SrTiO$_3$ substrates using a 90° off-axis sputtering technique [20]. We found that the $T_c$ of our SrRuO$_3$ film was ~150 K [21], a little less than the bulk value. Spectroscopic ellipsometry was used to acquire accurate and reproducible data of the complex dielectric function $\varepsilon(\omega)$ over a wide $T$ range. The spectroscopic ellipsometry unit was equipped with an ultra-high-vacuum cold-finger cryostat and operated in a photon energy range of 0.74 to 5.0 eV. For the density functional theory (DFT) calculations with LSDA, we used the full-potential linearized augmented plane-wave (FLAPW) band method [22], implemented in the WIEN2K package [23].

Figure 1(a) shows the $T$-dependent $\sigma(\omega)$ spectra of the SrRuO$_3$ film. As reported earlier [8, 24, 25], the optical transitions near 3 eV can be assigned to the charge transfer transition between the occupied O $2p$ band and the unoccupied Ru $t_{2g}$ band [8, 24]. The interband transition between Ru bands contributes a weak feature near $\hbar\omega \sim 1.5$ eV [24, 25]. As $T$ decreases, the interband transition peak near $\hbar\omega = 3.2$ eV gets significantly narrower, and the $\sigma(\omega)$ value near $\hbar\omega = 1.5$ eV for the Ru $d$–$d$ transition decreases. Note that the paramagnetic metal CaRuO$_3$ does not exhibit a such strong $T$ dependence in $\sigma(\omega)$, despite the fact that it has the same crystal symmetry and Ru valance as SrRuO$_3$, (see Fig. S1 in the Supplementary Materials [26]). This suggests that the sizeable spectral evolution in SrRuO$_3$ should be closely related to FM.

To obtain further evidence, we probed directly $\sigma(\Omega, T)$ at a fixed frequency of $\Omega$, while varying $T$. We measured the optical responses at four different photon energies while sweeping $T$ from 4 to 500 K in 2-K increment. Figures 1(b)–1(e) show the $T$ scans of $\sigma(\Omega, T)$ at $\hbar\Omega = 1.50$, 2.00, 2.63, and 3.27 eV. (As displayed later in Fig. 2(b), 2.63 and 3.27 eV correspond to the energies of the dip and peak structures, respectively, of the difference optical conductivity spectra.) All of the $T$-scan data exhibits clear kinks at $T_c$, indicating that there should exist a clear correlation between the FM order and $\sigma(\Omega, T)$.

In the Stoner model, the magnetic moment of the FM state is determined by the exchange splitting in the electronic structure. To understand how the change in magnetic moment affects $\sigma(\omega)$, we carried out LSDA calculations applying the fixed-spin-moment method [27]. Figure 2(a) shows the evolution of the calculated $\sigma(\omega)$ with various total spin polarizations ($m_{\text{tot}}$). The overall evolution in the calculated $\sigma(\omega)$ with increasing $m_{\text{tot}}$ agrees well qualitatively with that in the experimental $\sigma(\omega)$ with decreasing $T$. As $m_{\text{tot}}$ increases, the spectral weight of the interband $d$–$d$ transition below the 2.5-eV region decreases and two distinct peak structures at ~2.7 and 3.8 eV appear to merge. These features in the calculated $\sigma(\omega)$ should be related to the sharpening of the charge transfer peak structure in the experimental spectra (Fig. 1(a)). We obtained the difference spectra ($\Delta\sigma(\omega)$) by subtracting the high-$T$ data of $\sigma(\omega)$ at 450 K from the low-$T$ data of $\sigma(\omega)$ at 4 K. Although the two-peak structure could not be resolved in our experimental $\sigma(\omega)$, $\Delta\sigma(\omega)$ exhibits a dip at 2.7 eV and a peak at 3.3 eV. This structure is consistent with the calculated $\Delta\sigma(\omega)$, if the $\sigma(\omega)$ calculation with $m_{tot}$ =0.5 $\mu_B$ is subtracted from the $\sigma(\omega)$ calculation with $m_{tot}$ =1.9 $\mu_B$, as shown in Fig. 2(b).

For a better understanding of the spectral evolution, we took a closer look at the calculated band dispersions, shown in Figs. 2(c)–2(e). The Ru $t_{2g}$ bands are located between -1 and 1 eV, and the O $2p$ bands are located below -2eV. Figure 2(c) shows the band dispersions obtained from the non-spin-split band calculations, while Figs. 2(d) and 2(e) separately show minority and majority spin states obtained from the ferromagnetic band calculation. In the FM state, the energy levels of the Ru majority (minority) spin-split bands are lowered below (raised above) the Fermi level, which is consistent with the Stoner model. Meanwhile, the O $2p$ bands are hardly influenced by the spin polarization, as can be seen by comparing Fig. 2(c) with Figs. 2(d) and 2(e). The solid arrows correspond to the charge transfer transitions between occupied O $2p$ and unoccupied Ru $t_{2g}$ bands, supporting the earlier assignments [8, 24]. In contrast, the dashed arrows correspond to weak interband transitions between the Ru bands [24, 25].

According to Fermi's Golden Rule, the spectral weight of the optical transition should be proportional to the product of the occupied and unoccupied density of states (DOS) [28]. For the charge transfer transition, the DOS of the occupied O $2p$ bands will not change significantly, but that of the unoccupied Ru $t_{2g}$ bands will increase with $m_{tot}$. Thus its spectral weight change could be nearly proportional to $m_{tot}$. For the weak interband transition between Ru bands, the spectral weight change will depend on $\hbar\omega$. If $\hbar\omega$ is much larger than the exchange-splitting energy (~0.5 eV for SrRuO$_3$ estimated by band calculation [10, 25]), it will also be proportional to $m_{tot}$ (for more details, see Figs. S2 – and S3 in Supplementary Materials [26]). Because all of the $T$ scans of $\sigma(\Omega, T)$ were performed above 1.5 eV, their $\sigma(\Omega, T)$ should be nearly linear to $m_{tot}$. Therefore, the measured $T$-dependent evolution of $\sigma(\omega)$ and $\sigma(\Omega, T)$ could provide insight into how the local spin polarization evolves with $T$.

According to the original Stoner model [5], spin polarization is generated by the exchange splitting of band structures. Below $T_c$, spin polarization decays with $T^2$, due to the low-energy Stoner excitations between the electron–hole pairs of opposite spins [3, 29] as observed in some metallic ferromagnets such as $Ni_3Al$ and $ZrZn_2$ [30-32]. On the other hand, the model predicts that the spin polarization should vanish above $T_c$. A recent ARPES study showed that the FM exchange splitting remained nonzero above $T_c$, and suggested that the local moment FM was more appropriate for describing $SrRuO_3$ [9]. However, the $T$-dependent DOS changes could not be investigated in detail because it was difficult to obtain ARPES spectra in the high-$T$ region.

To highlight the $T$ evolution of the spin polarization in $SrRuO_3$, the $T$ scan $\sigma(\Omega, T)$ were further analyzed in detail. Below $T_c$, the change in $\sigma(\Omega, T)$ for all four measured frequencies deviated from linear behavior below $T_c$, as shown in Figs. 1(b)–(e), To find the $T$-dependence, for each $\hbar\Omega$ value, we normalized the change in $\sigma(\Omega, T)$ by the value $|\sigma(\Omega, T=T_c) - \sigma(\Omega, T=0 K)|$. All of the normalized $\sigma(\Omega, T)$ changes merged into one line and exhibits a $T^2$ dependence, independent of $\hbar\Omega$ (Fig. 3). This suggests that the local spin polarization change of $SrRuO_3$ should also have the same $T^2$ dependence. The $T^2$ behavior of our spectral changes is consistent with the predictions of the Stoner model, indicating that $SrRuO_3$ could be considered as a good candidate material for Stoner FM.

However, Figs. 1(b)–1(e) show that $\sigma(\Omega, T)$ still changes above $T_c$, which is inconsistent with the predictions of the original Stoner model. Taking a closer look above $T_c$, we can find that the $\sigma(\Omega, T)$ actually have an almost linear $T$ dependence. Note that this change of $\sigma(\Omega, T)$ can persist even at 500 K, which is more than 3 times larger than $T_c$. To obtain further insight, we looked at $\delta\sigma(\omega, T) = \sigma(\omega, T) - \sigma(\omega, T+50 K)$ with 50-K increments. Figure. 4(a) shows an increase in the overall intensity of $\delta\sigma(\omega, T)$ with increasing $T$ below $T_c$, while Fig. 4(b) exhibits a gradual decrease of overall intensity above $T_c$. Note that all of the $\delta\sigma(\omega, T)$ have very similar lineshapes, namely a dip around 2.6 eV and a peak around 3.2 eV. The similarity of the lineshapes of the $\delta\sigma(\omega, T)$ below and above $T_c$ suggests that $\delta\sigma(\omega, T)$ originates from same origin, namely the change in spin polarization. If the spin polarization vanished above $T_c$, then all of the $\delta\sigma(\omega, T)$ in Fig. 4(b) should have been essentially zero. Therefore, the non-vanishing features of $\delta\sigma(\omega, T)$ over the entire $T$ range signify that the spin polarization in the local band should change continuously with $T$, both below and even far above $T_c$. This implies that the expected energy shift of the spin split bands could be much smaller up to $T_c$, as the previous spectroscopic results implied [8, 9].

Persistent spin polarization above $T_c$ in SrRuO$_3$ has also been observed in other itinerant ferromagnetic systems such as Fe and Ni [33-35]. To explain the phenomenon, fluctuating local band theory of itinerant ferromagnetism has been developed [1, 4, 36], which is an extension of the original Stoner theory. Figures 5(a)–5(b) present schematics of prediction of local band theory for the paramagnetic phase just above $T_c$. Typical Heisenberg FM has collective spin excitations on a sub-micron length scale (which we denote by $\xi_m$) with a time scale of $10^{-13}$ s. In contrast, the itinerant FM is determined over a smaller length scale ($\xi_{lb}$ ~ several neighboring atoms) with a shorter time scale ($10^{-15}$ s) [2, 4]. In fluctuating local band theory, the long-range ferromagnetic order is destroyed at $T_c$ by the Heisenberg mechanism (Fig. 5(a)); whereas, the exchange splitting of band structures vanishes for a much higher $T$ scale by the Stoner mechanism (Figs. 5(b)–5(d)) [4, 36]. Therefore, just above $T_c$, the local band still has the exchange splitting as shown in Figs. 5(c) –5(d), but have a zero net magnetization due to spatial and temporal fluctuations.

The clear $T^2$ dependence below $T_c$ in our experimental $\sigma(\Omega, T)$ can be well explained by the original Stoner theory. However, there has been little research on the $T$ evolution of fluctuating local bands, even for the widely accepted Stoner systems, such as Fe, Co, and Ni. Ruban *et al* [37] performed Monte Carlo simulations based on *ab-initio* LSDA calculations, and showed that the magnitude of spin moment was nearly linear with respect to $T$ above $T_c$, consistent with our experimental observations. Further theoretical and experimental studies are needed to obtain detailed mechanism for the $T$-linear dependent spin polarization.

In summary, we provided a solid spectroscopic evidence of itinerant FM in $SrRuO_3$. The $T$ evolution observed in the $\sigma(\omega)$ spectra showed excellent agreement with the LSDA calculations with varying spin polarization. The $T^2$-scaling behavior in $\sigma(\Omega, T)$ below $T_c$ was also consistent with the Stoner model for itinerant FM. The non-vanishing local spin moments, even at higher $T$ (paramagnetic phase), were successfully interpreted by the fluctuating local band theory. Extensive studies on the $T$ evolution of fluctuating local bands will provide insight into the local versus itinerant magnetisms in spin-fluctuation-related physical phenomena, such as unconventional superconductivity and quantum criticality. Furthermore, our understanding on the nature of the itinerant FM will be important also in future device applications of oxide electronics, where $SrRuO_3$ is commonly used as an electrode material.

We acknowledge valuable discussions with J.G. Park, and Jaehyun Bae. This work was supported by the Research Center Program of IBS (Institute for Basic Science) in Korea. S.J.M. was supported by the Basic Science Research Program through the National Research Foundation (NRF) of Korea funded by the Ministry of Education, Science, and Technology (2012R1A1A1013274) and TJ Park Science Fellowship of POSCO TJ Park Foundation. J. H. S. was supported by Radiation Technology R&D program through NRF (2012029709). The work of at University of Wisconsin-Madison was supported by the AFOSR through grant FA9550-12-1-0342 and Army Research Office through grant W911NF-10-1-0362.

**References**

†soonjmoon@hanyang.ac.kr

*twnoh@snu.ac.kr

FIG 1. (color online). (a) Temperature ($T$) dependent optical conductivity, $\sigma(\omega)$, spectra of ferromagnetic SrRuO$_3$. (b)–(e) $T$ scans of $\sigma(\Omega, T)$ at $\hbar\Omega$ = 1.5, 2, 2.63, and 3.27 eV. All of the $T$ scans have clear kinks at $T_c$, suggesting strong correlation between the spectral evolution and the ferromagnetic order.

FIG 2. (color online). (a) Spin-polarization-dependent calculations of optical conductivity. The calculated $\sigma(\omega)$ demonstrates significant spectral weight redistribution, consistent with our experimental results. (b) Difference spectra taken from the experimental data ($\sigma(\omega)$ (450 K) – $\sigma(\omega)$ (5 K)) and from the theoretical calculations (LSDA; $\sigma(\omega)$ ($m_{tot}$ = 1.9$\mu_B$) – $\sigma(\omega)$ ($m_{tot}$ = 0.5$\mu_B$)). (c) The band dispersions obtained from the non-spin-split band calculation. (d)–(e) Band dispersions of the (d) minority and (e) majority spin states, obtained from spin-split band calculations. Ru $t_{2g}$ bands are located between -1 and 1eV, and O 2$p$ bands are located below -2 eV. The solid red and dashed blue arrows correspond to the 2.7 and 3.8 eV optical transitions in the calculated $\sigma(\omega)$, respectively.

FIG 3. (color online). $T$ scans of $\sigma(\Omega, T)$ for selected photon energies (1.5, 2, 2.63, and 3.27 eV). All of the normalized $\sigma(\Omega, T)$ data clearly demonstrate linearity with respect to $T^2$, consistent with the Stoner theory.

FIG 4. (color online). Difference spectra of the optical conductivities, $\delta\sigma(\omega,T) = \sigma(\omega,T) - \sigma(\omega,T+50$ K$)$, with a $T$ interval of $\delta T$ = 50 K (a) below $T = T_c$ and (b) above $T_c$. The gray dashed line represents the $\delta\sigma(\omega,T=400$K$)$ spectrum for comparison.

FIG 5. (color online). Schematic diagram of fluctuating local band theory. (a) A snapshot of the paramagnetic phase ($T > T_c$) on a macroscopic length scale ($\sim 1\mu m$). (b) Enlarged image of the black box in (a) for a length scale that shows slowly-varying collective spin fluctuation ($\xi_m \sim 0.1\ \mu m$). (c)–(d) Density of states of SrRuO$_3$, which indicates opposite local spin polarizations in the (c) green (upper) and (d) red (lower) box regions, respectively. Using optical

spectroscopy, we can probe the interband optical transition of local spin band structure within the local band length scale ($\xi_{lb}$).

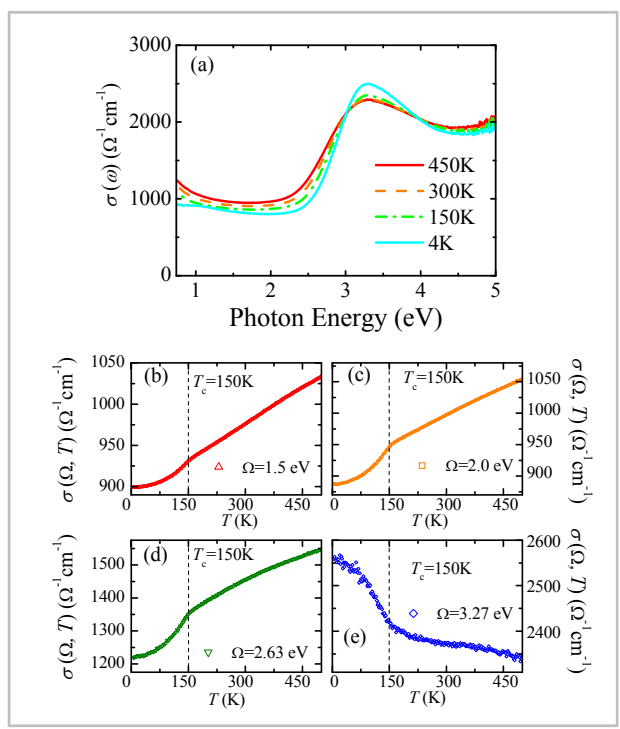

Figure. 1
Jeong *et al.*,

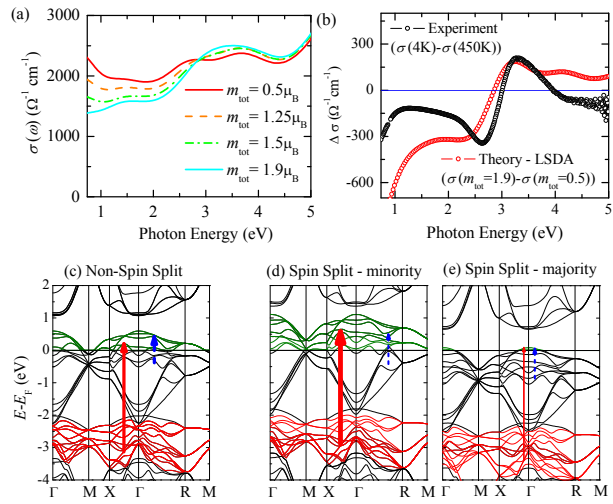

Figure. 2 Jeong et al.,

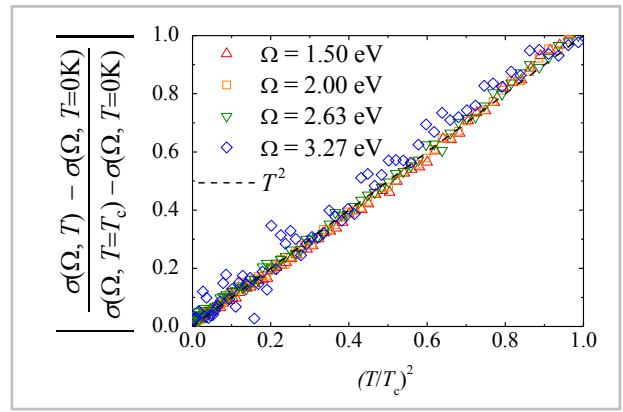

Figure. 3
Jeong *et al.*,

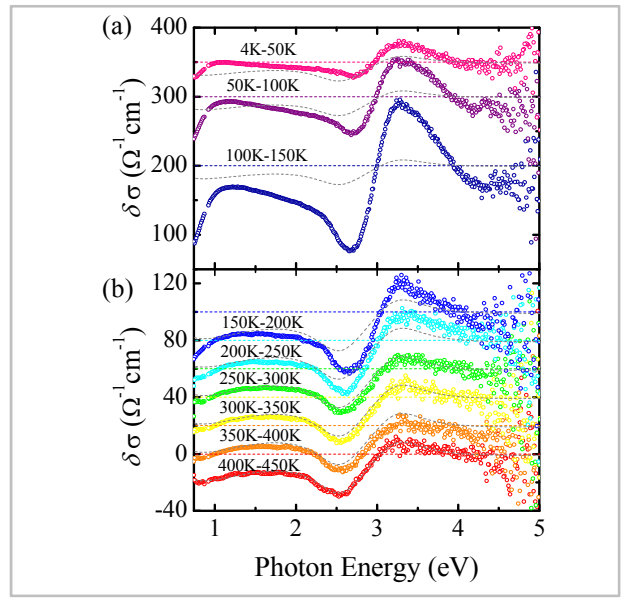

Figure. 4
Jeong *et al.*,

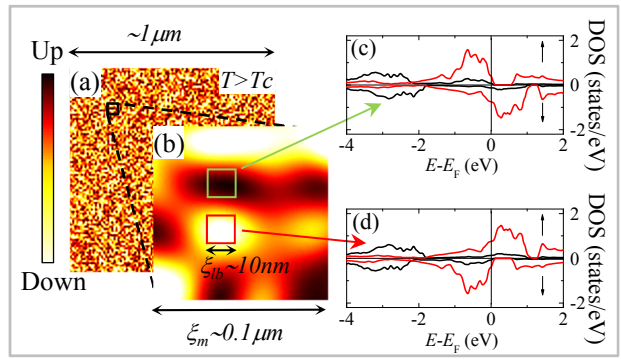

Figure. 5
Jeong *et al.*,